\documentclass[preprint,showpacs,preprintnumbers,amsmath,amssymb]{revtex4}
\usepackage[abs]{overpic}
\usepackage{graphicx,axodraw,subfigure}
\begin{document}
\preprint{HUPD1208}
\def\sla#1{\rlap/#1}
\def\tbr{\textcolor{red}}
\def\tcr{\textcolor{red}}
\def\ov{\overline}
\def\dprime{{\prime \prime}}
\def\nn{\nonumber}
\def\f{\frac}
\def\beq{\begin{equation}}
\def\eeq{\end{equation}}
\def\bea{\begin{eqnarray}}
\def\eea{\end{eqnarray}}
\def\bsub{\begin{subequations}}
\def\esub{\end{subequations}}
\def\dc{\stackrel{\leftrightarrow}{\partial}}
\def\ynu{y_{\nu}}
\def\ydu{y_{\triangle}}
\def\ynut{{y_{\nu}}^T}
\def\ynuv{y_{\nu}\frac{v}{\sqrt{2}}}
\def\ynuvt{{\ynut}\frac{v}{\sqrt{2}}}
\def\d{\partial}
\title{The pair production of Charged and Neutral Higgs bosons \\
in W and Z gauge boson fusion process}
\author{
%\footnote{},
Takuya  Morozumi
%\footnote{morozumi@hiroshima-u.ac.jp},
and   
Kotaro Tamai\\
}
\address{
Graduate School of Science, Hiroshima University,
Higashi-Hiroshima, 739-8526, Japan}
%\date{\today}
%%%%%%    TEXT START    %%%%%%
\def\nn{\nonumber}
\def\beq{\begin{equation}}
\def\eeq{\end{equation}}
\def\bei{\begin{itemize}}
\def\eei{\end{itemize}}
\def\bea{\begin{eqnarray}}
\def\eea{\end{eqnarray}}
\def\ynu{y_{\nu}}
\def\ydu{y_{\triangle}}
\def\ynut{{y_{\nu}}^T}
\def\ynuv{y_{\nu}\frac{v}{\sqrt{2}}}
\def\ynuvt{{\ynut}\frac{v}{\sqrt{2}}}
\def\s{\partial \hspace{-.47em}/}
\def\ad{\overleftrightarrow{\partial}}
\def\ss{s \hspace{-.47em}/}
\def\pp{p \hspace{-.47em}/}
\def\bos{\boldsymbol}
\begin{abstract}
We study the signatures of a two Higgs doublet model of Davidson and Logan.
The model includes an extra Higgs doublet 
with the vacuum expectation value (VEV) much smaller than the one of the
standard model like Higgs. The smaller VEV is related to the origin of the
small neutrino mass in the two Higgs doublet model.
In the model, a single non-standard model like Higgs 
production of weak gauge boson fusion is suppressed due to the
smallness of the vacuum expectation value.  In contrast to the single Higgs production,
the cross section 
of the Higgs pair production due to gauge boson fusion
is not suppressed.  Using the model, we compute
the charged Higgs and neutral Higgs pair production cross section in  
W$^+$ Z annihilation channel.
In the two Higgs doublet model, the charged Higgs $H^+$
decays into a pair of the charged anti-lepton and right-handed neutrino 
. The neutral Higgs boson decays into right-handed neutrino and left-handed
anti-neutrino pair which is invisible.
A single charged anti-lepton and three neutrinos are the products
of the subsequent decays of the charged Higgs and the neutral Higgs.
W$^+$ Z pair production gives the background for the signal
through the decays $W^+ \to \nu \bar{l}$ and $Z \to \nu \bar{\nu}$. 
By multiplying the charged and neutral Higgses production cross section with 
the lepton flavor specific decay branching fractions of charged Higgs, 
we define a measurement which characterizes the present model.
We numerically compute the measurement and find the sizable deviation
from the standard model prediction.
\end{abstract}
\maketitle
%%%%%%    TEXT START    %%%%%%
\section{Introduction}
Probing the Higgs sector beyond the standard model is a main subject of 
new physics search in collider experiments and in flavor
factories. Among them, two Higgs doublet models have been studied
in their many aspects.  Here we study the two Higgs doublet model
in which the hierarchy of the neutrino mass and the other 
fermions mass is explained by two Higgs vacuum expectation values (VEVs)
with large hierarchy \cite{Davidson:2009ha,Davidson:2010sf}.
In the model, the neutrino mass is protected 
from the large standard model like Higgs VEV by assigning U(1) charge for 
right-handed neutrino
and the second Higgs. The corresponding U(1) symmetry is softly broken.
Moreover the tiny VEV of the second Higgs is stable against the 
radiative
corrections, \cite{Morozumi:2011zu, Haba:2010zi}.
   In this letter, we study the production of the 
Higgs bosons in the second doublet in 
gauge bosons fusion process.  A single Higgs boson production of gauge boson fusion
is suppressed because the amplitude is proportional to the smaller Higgs 
VEV. In contrast to the single Higgs boson production,
the Higgs boson pair production of gauge boson fusion is not suppressed. 
Therefore it is worthwhile to study the process in their detail. We study the 
pair production cross section of the charged Higgs and the neutral Higgs
in W and Z boson fusion process.  Since the neutral Higgs bosons decay into neutrino
and anti-neutrino pair, they are invisible. Therefore the signal of the event
is a single charged lepton of the charged Higgs decay.  The standard model
background of the process is related to W Z scattering process. Invisible decay of Z boson 
with the W decay into lepton and neutrino pair is a potential background.
In this letter, we compute the differential cross section of the
charged and neutral Higgses pair production cross section and compare
it with W$^+$ Z pair production in the standard model.
Since the charged Higgs coupling to charged lepton depends on the
lepton flavor, we expect the strong flavor dependence
in the charged Higgs decay while W decay is flavor independent.

The paper is organized as follows. In section (II),
we study the single Higgs production.  In section (III), the Higgs pair production
cross section and the angular distribution 
of the production cross section are studied and they are compared with the one
of $W+Z$ scattering. Section (IV), we propose a measurement which identifies
the signal and predictions are given.  A brief summary is also given.
\section{A single Higgs production from weak gauge boson fusion}
We study the Dirac neutrino model of \cite{Davidson:2009ha,
Davidson:2010sf}. In this two Higgs
doublet model, besides the standard model like Higgs, there are
two extra neutral Higgs and one charged Higgs. We follow the notation 
of \cite{Morozumi:2011zu}.

We first consider the
single Higgs production with the gauge boson fusion.
Denoting the standard model like Higgs as H, and the other two
neutral Higges as A(CP odd Higgs) and h(CP even Higgs) respectively,
the interaction Hamiltonian for gauge boson fusion to neutral Higgses
is given as,  
\bea
{\cal L}=gM_W (W_\mu^+ W^{\mu -}+\frac{1}{2 c_W^2} Z^\mu Z_\mu)
\left(\sin(\beta+\gamma) h + \cos(\beta +\gamma) H \right),
\label{eq:singlehiggs}
\eea 
where $\tan \beta$ is the ratio of the two vacuum expectation values
of Higgs and is given as,
\bea
\frac{v_2}{v_1}=\tan \beta \simeq \frac{m_{12}^2}{M_A^2},
\eea
where $m_{12}^2$ is the soft breaking term for U(1) symmetry.
Since $v_1$ is standard model like Higgs VEV and $v_2$ is VEV of the Higgs
which gives rise to the light neutrino
masses, $\tan \beta$ is very small. If one takes $v_2= 1$(eV), 
then $tan \beta=O(10^{-11})$.
$\gamma$ is related to a mixing angle of CP even Higgses. 
Since the ratio of $\frac{\gamma}{\beta}$ is given as
$
\frac{\gamma}{\beta} 
\simeq \frac{M_A^2-\frac{\lambda_3+\lambda_4}{\lambda_1}M_H^2}{M_H^2-M_A^2}
$,  $\gamma$ is as small as $\beta$.  
%Therefore the single non-standard model like 
%Higgs ($h$) production due to gauge boson fusion is suppressed.  
From Eq.(\ref{eq:singlehiggs}), we note only the CP even Higges h and H can be
produced by the gauge bosons fusion. 
Since the coupling of the non-standard like Higgs h with gauge boson pairs
is proportional to $\sin (\beta + \gamma)$, 
the single non-standard model like Higgs
production is strongly suppressed. 
\section{Higgs pair production from weak gauge boson fusion}
In contrast to the single Higgs production,
the pair production of the extra Higgses is not suppressed by the
small VEV. Here we consider the charged Higgs ($H^+$) and a neutral Higgs
($h$ or $A$) pair production due to $W^+$ $Z$ fusion.  
In the neutrinophilic model \cite{Davidson:2009ha}-\cite{Haba:2010zi}, the non-standard model like Higg couples to
left-handed lepton doublets and right-handed singlet neutrinos.  
The visible
decay product of the pair produced Higgses is a mono charged lepton jet.
The other decay products are three (anti-)neutrinos which are invisible.
We first study the cross section of $W^+ + Z \to H^+ + h$
and $W^+ + Z \to H^+ + A $. 
The interaction vertices for the pair production are,
\bea
{\cal L}&=& \frac{g^2}{2} s_W (A_\mu -t_W Z_\mu)[(H^+ W^{\mu -}+H^- W^{\mu +})
(h \cos(\beta+\gamma)-H \sin(\beta+\gamma))\nn \\
&-&i(H^+ W^{\mu -}-H^- W^{\mu +})
A] \nn \\
&+& i \frac{g \cos 2 \theta_W }{2 \cos \theta_W} Z_\mu (\partial^\mu H^- H^+-
\partial^\mu H^+ H^-) \nn \\
&+& \frac{g \cos(\beta+\gamma)}{2 \cos \theta_W}
(\partial_\mu h A- \partial_\mu A h) Z^\mu \nn \\
&+&\Bigl{\{} i \frac{g}{2} \cos(\beta+\gamma) W^{+ \mu}(h \partial_\mu H^-- \partial_\mu h H^-)
+\frac{g}{2} 
W^{+\mu}(H^-\partial_\mu A- A \partial_\mu H^-) + h.c. \Bigr{\}}.
\label{eq:int}
\eea
Then the amplitude for $Z + W^+ \to h + H^+$ is,
\bea
M&=&\epsilon_W^\mu \epsilon_Z^\nu T_{\mu \nu}, \nn \\ 
T_{\mu \nu}&=& \frac{g^2 \cos(\beta+\gamma)}{2\cos \theta_W}
\left(a g_{\mu \nu}+ d q_{h \nu} q_{H^+ \mu}
+b  q_{H^+ \nu} q_{h \mu}\right),
\label{eq:hamp}
\eea 
where we compute the four Feynman diagrams corresponding to the s channel $W^+$
exchange(Fig.~1), the contact interaction (Fig.~2), 
u channel charged Higgs exchange (Fig.~3), and 
t channel CP odd Higgs (A) exchanged diagram (Fig.~4).
$a, b$ and $d$ in Eq.(\ref{eq:hamp}) are given as,
\bea
a&=&-s_W^2-s_W^2\frac{M_h^2-M_{H^+}^2-M_W^2}{s-M_W^2}-c_W^2
\frac{t-u+M_Z^2-M_W^2}{s-M_W^2}, \nn \\
b&=&\frac{2\cos 2 \theta_W}{u-M_{H^+}^2}+ \frac{2(\cos 2 \theta_W+1)}{s-M_W^2},\nn \\
d&=&-\frac{2}{t-M_A^2}-\frac{2(\cos 2 \theta_W+1)}{s-M_W^2},
\eea
with $s_W^2=\sin^2 \theta_W, c_W^2=\cos^2 \theta_W, t=(q_{H^+}-p_W)^2$,
and $u=(p_W-q_h)^2$. $t$ and $u$ 
are also written in terms of the center of mass (CM) energy 
$\sqrt{s}$ and the scattering angle $\theta$ between $W^+$ and $H^+$ in CM frame
as,
\bea
t&=&M_{H^+}^2+M_{W^+}^2-\frac{(s+M_W^2-M_Z^2)(s+M_{H^+}^2-M_h^2)}{2s}+ 2 P_{W^+} P_{h} 
\cos \theta, \nn \\
u&=&-s+M_{Z}^2+M_{h}^2+\frac{(s+M_W^2-M_Z^2)(s+M_{H^+}^2-M_h^2)}{2s}-2 P_{W^+} P_{h}
\cos \theta.
\label{eq:tu}
\eea 
Using Eq.(\ref{eq:int}), one can compute the cross section.
The unpolarized cross section of $W^+$ and $ Z$ fusion into the charged Higgs $(H^+)$
and  neutral Higgs $(h)$pair is given as,
\bea
&&\frac{d \sigma_{W^+ + Z \rightarrow H^+ +h}}{d \cos \theta}
=\frac{g^4 \cos(\beta+\gamma)^2}{36 c_W^2}
\frac{1}{32 \pi s}
\frac{P_{h}}{P_{W^+}} \nn \\
&& 
\Biggr{[}a^2\Bigr\{3+ \frac{(s-(M_{W^+}+M_Z)^2)(s-(M_{W^+}-M_Z)^2)}{4 M_Z^2 M_W^2}\Bigl\}\nn \\
&&+b^2
\Bigr\{\left(\frac{(M_{h}^2+M_W^2-u)^2}{4 M_W^2}-M_{h}^2 \right)
\left(\frac{(M_{H^+}^2+M_Z^2-u)^2}{4 M_Z^2}-M_{H^+}^2\right)
\Bigl\} \nn \\
&&+d^2 
\Bigr\{\left(\frac{(M_{H^+}^2+M_W^2-t)^2}{4 M_W^2}-M_{H^+}^2 \right)
\left(\frac{(M_{h}^2+M_Z^2-t)^2}{4 M_Z^2}-M_{h}^2\right)
\Bigl\} \nn \\
&&+2 b d
 \left(\frac{s-M_h^2-M_{H^+}^2}{2}-
\frac{(M_h^2+M_W^2-u)(M_{H^+}^2+M_W^2-t)}{4 M_W^2}
\right) 
\nn \\
&& \times \left(\frac{s-M_h^2-M_{H^+}^2}{2}-
\frac{(M_h^2+M_Z^2-t)(M_{H^+}^2+M_Z^2-u)}{4 M_Z^2}
\right) \nn \\
&&+a d \Bigl\{-
\frac{(M_h^2+M_Z^2-t)(M_{H^+}^2+M_W^2-t)(M_W^2+M_Z^2-s)}{4 M_W^2 M_Z^2}
+s-M_{H^+}^2-M_h^2 \nn \\
&&-\frac{(M_h^2+M_W^2-u)(M_{H^+}^2+M_W^2-t)}{2 M_W^2}
-\frac{(M_h^2+M_Z^2-t)(M_{H^+}^2+M_Z^2-u)}{2 M_Z^2}
 \Bigr\} \nn \\
&&+a b \Bigl\{-
\frac{(M_h^2+M_W^2-u)(M_{H^+}^2+M_Z^2-u)(M_W^2+M_Z^2-s)}{4 M_W^2 M_Z^2}
+s-M_{H^+}^2-M_h^2 \nn \\
&&-\frac{(M_h^2+M_W^2-u)(M_{H^+}^2+M_W^2-t)}{2 M_W^2}
-\frac{(M_h^2+M_Z^2-t)(M_{H^+}^2+M_Z^2-u)}{2 M_Z^2}
 \Bigr\}
\Biggl{]},
\eea
$P_{W^+}$ and $P_{h}$ are momentum of $W^+$ and $h$ in the CM frame,
\bea
P_{h}&=& \frac{\sqrt{
(s^2-2s(M_{H^+}^2+M_h^2)+(M_{H^+}^2-M_h^2)^2)}}{2 \sqrt{s}},\nn \\
P_{W^+}&=& \frac{\sqrt{(s^2-2s(M_W^2+M_Z^2)+(M_W^2-M_Z^2)^2)}}{2 \sqrt{s}}.
\eea
\begin{figure}[htb]
\begin{center}
\includegraphics[scale=0.6]{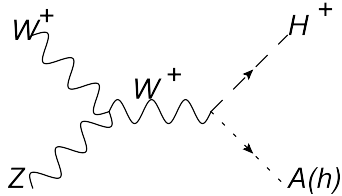}
\caption{S channel W exchange}
\end{center}
\end{figure}
\begin{figure}[htb]
\begin{tabular}{ccc}
\begin{minipage}{0.28\hsize}
\begin{center}
\includegraphics[scale=0.5]{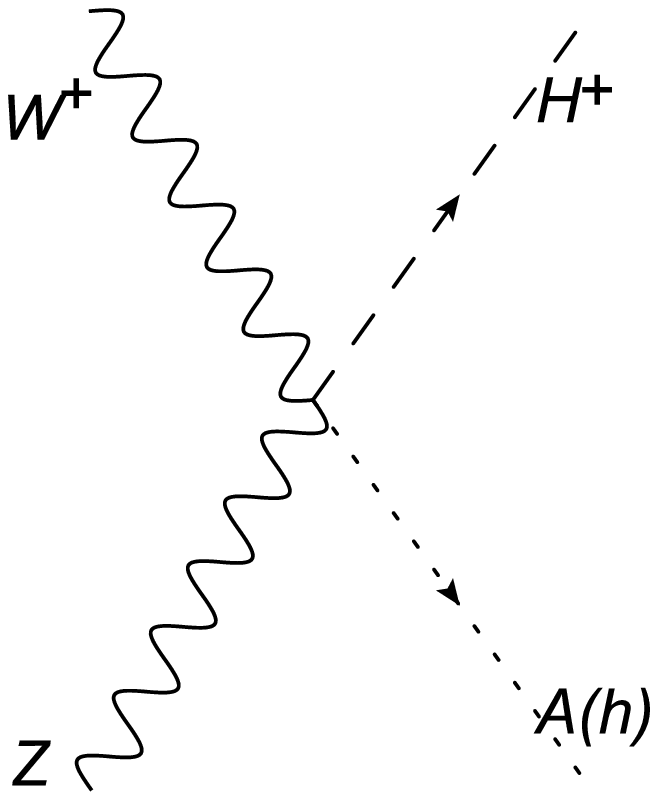}
\caption{Contact interaction}
\end{center}
\end{minipage}
\begin{minipage}{0.28\hsize}
\begin{center}
\includegraphics[scale=0.5]{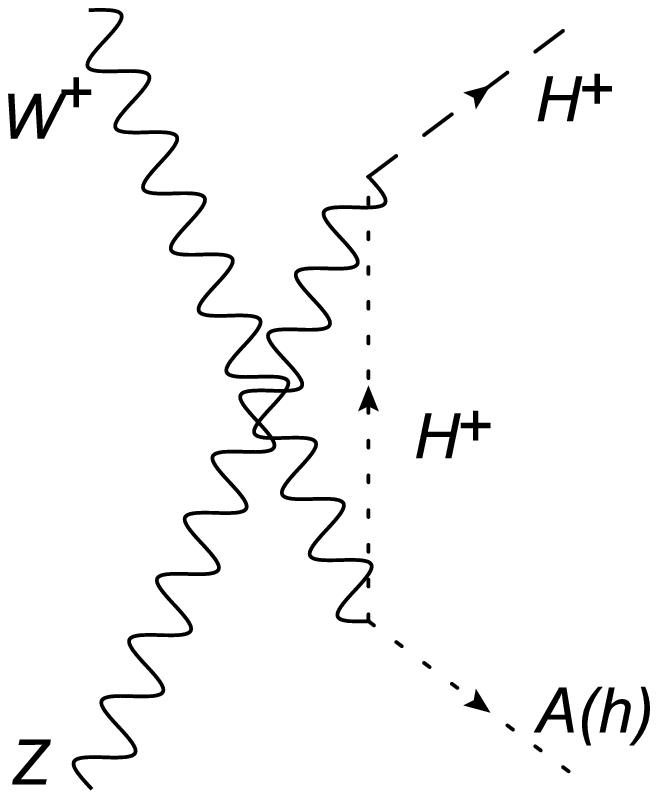}
\caption{U channel}
\end{center}
\end{minipage}
\begin{minipage}{0.28\hsize}
\begin{center}
\includegraphics[scale=0.5]{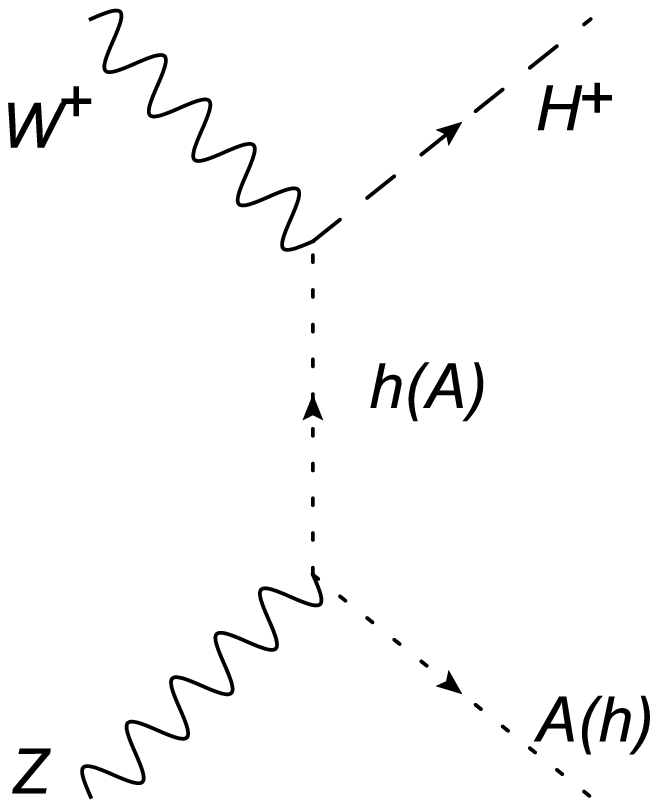}
\caption{T channel}
\end{center}
\end{minipage}
\end{tabular}
\end{figure}
In the same way as that of the CP even Higgs production,
the amplitude for CP odd Higgs and charged Higgs pair production
is obtained by computing the Feynman diagrams corresponding to Figs.~1-4.
\bea
M_{A}&=&\epsilon_W^\mu \epsilon_Z^\nu T_{A \mu \nu}, \nn \\ 
T_{A \mu \nu}&=& \frac{-i g^2}{2\cos \theta_W}
\left(a_A g_{\mu \nu}+ d_A q_{A \nu} q_{H^+ \mu}
+b_A  q_{H^+ \nu} q_{A \mu}\right).
\label{eq:Aamp}
\eea 
$a_A, b_A$ and $d_A$ in Eq.(\ref{eq:Aamp}) are given as,
\bea
a_A&=&s_W^2+s_W^2\frac{M_A^2-M_{H^+}^2-M_W^2}{s-M_W^2}+c_W^2
\frac{t_A-u_A+M_Z^2-M_W^2}{s-M_W^2}, \nn \\
b_A&=&-\frac{2\cos 2 \theta_W}{u_A-M_{H^+}^2}- \frac{2(\cos 2 \theta_W+1)}{s-M_W^2},\nn \\
d_A&=&\frac{2}{t_A-M_h^2}+\frac{2(\cos 2 \theta_W+1)}{s-M_W^2},
\eea
with $t_A=(q_{H^+}-p_W)^2, u_A=(p_W-q_A)^2$.
The kinematical variables $t_A$ and $u_A$ for $H^+$ and $A$ production can be obtained by
replacing $M_h$ by $M_A$ in the Mandelstam variables $t,u$ in Eq.(\ref{eq:tu}),  
\bea
t_A&=&M_{H^+}^2+M_{W^+}^2-\frac{(s+M_W^2-M_Z^2)(s+M_{H^+}^2-M_A^2)}{2s}+ 2 P_{W^+} P_{A}
 \cos \theta, \nn \\
u_A&=&-s+M_{Z}^2+M_{A}^2+\frac{(s+M_W^2-M_Z^2)(s+M_{H^+}^2-M_A^2)}{2s}-
2 P_{W^+} P_{A} \cos \theta.
\eea
Since $\beta$ and $\gamma$ is very small and CP even and CP odd
Higgses are almost degenerate, $M_h \simeq M_A$, the production cross sections
for $H^+ + h$ and $H^+ + A$ are almost identical. 
Then the production cross section for charged
Higgs and CP odd Higgs $(A)$ is given as,
\bea
&& \frac{ d\sigma_{W^+ + Z \rightarrow H^+ + A}}{ d \cos \theta}=
\frac{g^4}{36 c_W^2}
\frac{1}{32 \pi s}\frac{P_A}{P_W^+} \nn \\
&& 
\Biggr{[}a_A^2\Bigr\{3+ \frac{(s-(M_{W^+}+M_Z)^2)(s-(M_{W^+}-M_Z)^2)}{4 M_Z^2 M_W^2}\Bigl\}\nn \\
&&+b_A^2
\Bigr\{\left(\frac{(M_{A}^2+M_W^2-u_A)^2}{4 M_W^2}-M_{A}^2 \right)
\left(\frac{(M_{H^+}^2+M_Z^2-u_A)^2}{4 M_Z^2}-M_{H^+}^2\right)
\Bigl\} \nn \\
&&+d_A^2 
\Bigr\{\left(\frac{(M_{H^+}^2+M_W^2-t_A)^2}{4 M_W^2}-M_{H^+}^2 \right)
\left(\frac{(M_{A}^2+M_Z^2-t_A)^2}{4 M_Z^2}-M_{A}^2\right)
\Bigl\} \nn \\
&&+b_A d_A
\frac{
M_A^2(M_{H^+}^2+3 M_W^2-t_A)+2M_W^2(M_{H^+}^2-s)+(M_{H^+}^2+M_W^2-t_A)(M_W^2-u_A)
}{8M_W^2 M_Z^2} \times \nn \\
&&\{M_A^2 (M_{H^+}^2 + 3 M_Z^2 - u_A) + M_{H^+}^2 (3 M_Z^2 - t_A) + M_Z^4 - 
2 M_Z^2 s - 
  M_Z^2 t_A - M_Z^2 u_A + t_A u_A)\} \nn \\
&&+a_A d_A \Bigl\{-
\frac{(M_A^2+M_Z^2-t_A)(M_{H^+}^2+M_W^2-t_A)(M_W^2+M_Z^2-s)}{4 M_W^2 M_Z^2}
+s-M_{H^+}^2-M_A^2 \nn \\
&&-\frac{(M_A^2+M_W^2-u_A)(M_{H^+}^2+M_W^2-t_A)}{2 M_W^2}
-\frac{(M_A^2+M_Z^2-t_A)(M_{H^+}^2+M_Z^2-u_A)}{2 M_Z^2}
 \Bigr\} \nn \\
&&+a_A b_A \Bigl\{-
\frac{(M_A^2+M_W^2-u_A)(M_{H^+}^2+M_Z^2-u_A)(M_W^2+M_Z^2-s)}{4 M_W^2 M_Z^2}
+s-M_{H^+}^2-M_A^2 \nn \\
&&-\frac{(M_A^2+M_W^2-u_A)(M_{H^+}^2+M_W^2-t_A)}{2 M_W^2}
-\frac{(M_A^2+M_Z^2-t_A)(M_{H^+}^2+M_Z^2-u_A)}{2 M_Z^2}
 \Bigr\}
\Biggl{]}.
\eea
%%%%%%%%%%%%%%%%%%%%%%%%%%Numerical stuff%%%%%%%%%%%%%%%%%%%%%%%%%%%%%%%%%%%%%%%%%%%%%%
In Fig.~5, we show the differential cross section 
$\frac{d \sigma}{d \cos \theta}\Big{|}_{\sqrt{s}=600({\rm GeV})}$
for the neutral Higgs ($h$ or $A$) and charged Higgs pair production.
The numerical values for $(M_W, M_Z, \sin^2 \theta_W)$ are
$(80({\rm GeV}),91({\rm GeV}),0.23)$.
We show the cross sections for three cases for Higgs spectrum in GeV unit as
$(M_X, M_{H^+})=(200, 200),(200, 300)$, and $(300, 200)$ with $X=A$ or $h$. 
The differential cross sections for non-degenerate masses of charged Higgs and neutral Higgs 
is much larger than the case
that they are degenerate.
We also study the dependence of the total cross section on CM energy $\sqrt{s}$
in Fig.~6.
The total cross section decreases as $\sqrt{s}$ increases.
For degenerate case,
the total cross section at $\sqrt{s}=600$ (GeV) is about 0.02 pb. 
When they are non-degenerate, the cross section is much  
larger than that of the degenerate case as shown in Fig.~6. 
The cross section for non-degenerate case is as large as 0.8 pb at $\sqrt{s}=600$ (GeV). 
\begin{figure}[htb]
\begin{center}
\includegraphics[scale=0.7]{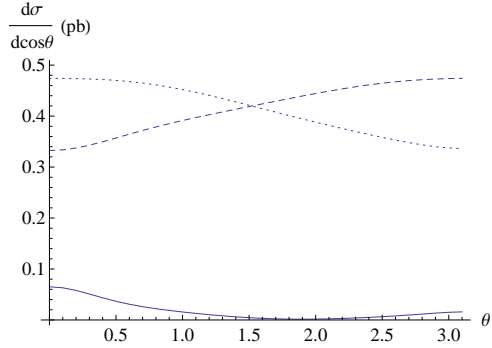}
\caption{The production cross section $\frac{d \sigma}{ d \cos \theta}$ of 
a pair of neutral Higgs $X=A,h$ and charged Higgs $H^+$ for $\sqrt{s}=600$(GeV).
The solid line corresponds to $(M_X,M_{H^+})=(200,200)$(GeV). The dashed line corresponds 
to $(M_X,M_{H^+})=(200,300)$(GeV). 
The dotted line corresponds to $(M_X,M_{H^+})=(300,200)$(GeV).}
\end{center}
\end{figure}
\begin{figure}[htb]
\begin{center}
\includegraphics[scale=0.8]{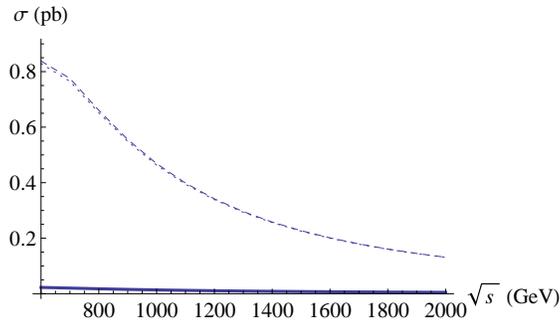}
\caption{ $\sigma(W+Z \to H^+ + X)$ ($X=A,h$) as a function of $\sqrt{s}$. The dashed 
line  corresponds to $(M_X,M_{H^+})=(200,300)$(GeV). The dotted line corresponds to
$(M_X,M_{H^+})=(300,200)$ (GeV). The solid line corresponds to $(M_X, M_{H^+})=
(200,200)$ (GeV).} 
\end{center}
\end{figure}
%%%%%%%%%%%%%%%%%%%%%%%%%%%%%%%%%%%%%%%%%%%%%%%%%%%%%%%%%%%%%%%%%%%%%%%%%
\section{Signal of the pair production of the neutral Higgs and charged Higgs,
and standard model background}
The signals of the neutral Higgs production associated with a single 
charged Higgs production are large missing momentum 
and energies. The charged Higgs boson can be identified from their decays into 
a pair of charged lepton and neutrino. On the other hand, the neutral Higgs A or h decays
into a pair of neutrino and antineutrino.
Therefore, one can not see the decay products of 
neutral Higgs which leads 
to the large missing momentum in CM frame of the vector bosons fusion.
The signal 
for the charged Higgs and the neutral Higgs pair production
of the present model is the single charged lepton from the charged Higgs decay. The
signal event 
occurs through the subsequent decays of the charged Higgs and the neutral Higgs
into $l^+ \nu$
and $\bar{\nu} \nu$ respectively.
Within the standard model, the event which is similar to the signal 
can occur through the process,
$W^+ + Z  \to W^+ + Z \to l^+ \nu_l \bar{\nu_k} \nu_k
$.
In the following, we propose a measurement which characterizes 
the present model and estimate it using the production cross section 
computed in the previous section.
We define the ratio of 
the numbers of the two events, one is the number of events with 
a single charged lepton with three neutrinos 
and the other is the number of the events with 
a single charged lepton and a $b \bar{b}$ pair,
\bea
R_l=\frac{\sum_{k} N(W^+ + Z \to \nu_k \bar{\nu}_k l^+ \nu_l)}
{N(W^+ + Z \to b \bar{b} l^+ \nu_l)},
\eea
where the charged lepton flavor is specified as $l$. 
One can write the ratio as,
\bea
R_l=\sum_{X=h,A} \left(\frac{\sigma(W^+ + Z \to H^+ +X)}{\sigma^{SM}(W^+ + Z \to W^+ + Z)}\frac{Br(X \to \nu \bar{\nu})}{Br(Z \to b \bar{b})} \right)
\frac{Br(H^+ \to l^+ \nu_l)}{Br(W^+ \to l^+ \nu_l)}+
\frac{Br(Z \to \nu \bar{\nu})}{Br(Z \to b \bar{b})},\nn \\
\label{eq:ratio}
\eea
where we used the shorthand notation,
$
Br(X \to \nu \bar{\nu})=\sum_{k}Br( X \to \nu_k \bar{\nu}_k)
$ 
for $X=h,A,Z$.
Below we compute $R_l$.
To compute $R_l$,  one needs the ratio of the cross 
sections  for Higgs pair production and vector boson pair production. The cross section of W Z scattering is calculated \cite{Bahnik:1997ka, Green:2003if} and the ratio of the cross section,
\bea
\frac{2 \sigma(W^+ + Z \to H^+ + X)}{\sigma_{SM}(W^+ + Z \to W^+ + Z )},
\label{eq:cratio}
\eea
is shown in Fig.~7.
Since Higgs pair production cross section is at most 0.8 pb while gauge bosons
scattering cross section is about 200 pb, the ratio can be as large as 
a few times $10^{-3}$. In what follows, we use $6.0 \times 10^{-3}$
as a numerical value for the ratio of the cross sections of Eq.(\ref{eq:cratio}).
The other branching fractions which appear
in Eq.(\ref{eq:ratio}) are quoted from Particle Data Group (PDG) \cite{PDG:2012},
\bea
Br(W^+ \to \tau^+ \nu)&=&11.25 \pm 0.20 \%, \\
Br(W^+ \to \mu^+ \nu)&=&10.57 \pm 0.15 \%, \\
Br(W^+ \to e^+ \nu)&=&10.75 \pm 0.13 \%, \\
Br(Z \to b \bar{b})&=&15.12 \pm 0.05 \%, \\
Br(Z \to \nu \bar{\nu})&=& 20.00 \pm 0.06 \%.
\eea
For the new physics side, the neutral Higgses can only decay into
neutrino and anti-neutrino pairs. Therefore, 
$Br(h \to \nu \bar{\nu})=Br(A \to \nu \bar{\nu})=100 \%$.
Using the numerical values, one can write $R_l (l=e, \mu,\tau) $ as,
\bea
R_e&=&1.32+0.61 \times
Br(H^+ \to e^+ \nu)\frac{2 \sigma(W^+ + Z \to H^+ +X) }{\sigma^{SM}(W^+ + Z \to W^+ + Z)}
,\nn \\
R_\mu&=&1.32+0.63 \times
Br(H^+ \to \mu^+ \nu)\frac{2 \sigma(W^+ + Z \to H^+ +X) }{\sigma^{SM}(W^+ + Z \to W^+ + Z)}
,\nn \\
R_\tau&=&1.32+0.59 \times
Br(H^+ \to \tau^+ \nu)\frac{2 \sigma(W^+ + Z \to H^+ +X) }{\sigma^{SM}(W^+ + Z \to W^+ + Z)},
\eea
where $Br(H^+ \to l \nu)$ in $\%$ unit should be substituted.
\begin{figure}
\begin{center}
\includegraphics[scale=0.8]{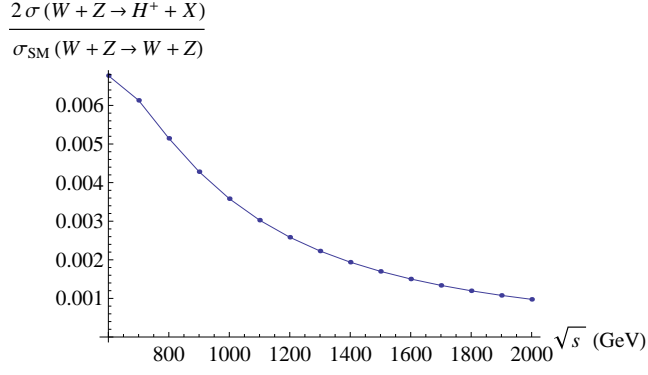}
\end{center}
\caption{The ratio of the cross sections 
$\frac{2 \sigma(W + Z \to H^+ + X)}{\sigma_{\rm SM}(W^+ + Z \to W^+ Z)}$
as a function of $\sqrt{s}$(GeV) for non-degenerate charged Higgs and neutral Higgs.
We choose their masses as $M_{H^+}=300$(GeV), $M_{X}=200$(GeV) $X=h,A$.}
\end{figure}
The charged Higgs can decay into charged leptons and neutrino.
In contrast to the leptonic decay of W boson, the branching fractions 
are dependent on the lepton flavor as, 
$
Br(H^+ \to l^+ \nu_l)=\frac{\sum_{i=1}^3 {m_i^2 } |V_{li}|^2}{\sum_{i=1}^3 
m_i^2}$ \cite{Davidson:2009ha}.
We update the branching fraction to each lepton flavor mode
using the recent results on $|V_{e3}|$.
For normal hierarchy case, the branching fractions are written as,
\bea
{\rm Br}(H^+ \to l^+ \nu_l)=\frac{ m_1^2+\Delta m_{sol.}^2 |V_{l2}|^2
+(\Delta m_{sol}^2 + \Delta m_{atm}^2) |V_{l3}|^2}{3 m_1^2
+ 2\Delta m_{sol}^2 + \Delta m_{atm}^2}.
\label{eq:normal}
\eea
In the formulae of Eq.(\ref{eq:normal}), $m_1$ denotes the lightest neutrino mass.
For inverted hierarchical case, they are written as,
\bea
{\rm Br}(H^+ \to l^+ \nu_l)=\frac{ m_3^2+\Delta m_{atm.}^2 
(|V_{l1}|^2+|V_{l2}|^2)
-\Delta m_{sol}^2 |V_{l1}|^2}{3 m_3^2
+ 2\Delta m_{atm}^2 - \Delta m_{sol}^2},
\eea
where $m_3$ denotes the lightest neutrino mass.
We have used the following value for the mixing angles \cite{PDG:2012},
$
\sin^2 \theta_{12}=0.306, \ 
\sin^2 \theta_{23}=0.42, \ 
\sin^2 \theta_{13}=0.021.
$
We also use $m_{atm}^2=2.35 \times 10^{-3}$(eV$^2$) and 
$ m_{sol}^2=7.58 \times 10^{-5}$ (eV$^2$) for mass squared 
differences for neutrino mass eigenstates.
In Fig.~8, we have shown $R_l$ ($l=e, \mu, \tau)$ for normal hierarchical case as functions of
the lightest neutrino mass $m_1$. In Fig.~9, we have shown $R_l$ for inverted hierarchycal
case as functions for the lightest neutrino mass $m_3$. The shaded bands in Figs.~8,9,
show the standard model expectation and errors given as $1.322\pm 0.006$.
\begin{figure}[htb]
\begin{center}
\begin{tabular}{lcr}
\begin{minipage}{0.4\hsize}
\begin{center}
\includegraphics[width=6cm]{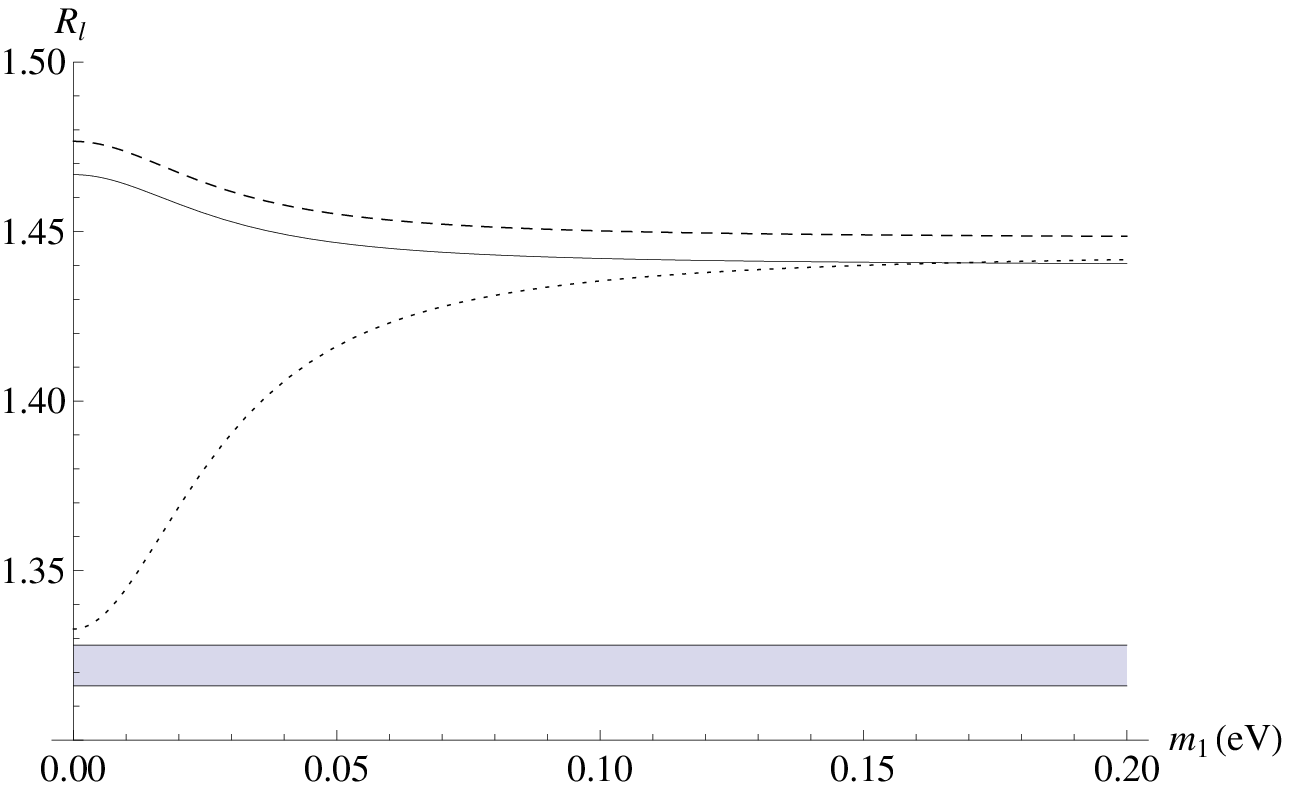}
\caption{ $R_l$ $(l=e, \mu, \tau)$ 
for normal hierarchical case as  functions of the lightest neutrino mass $m_1$(eV). 
The dotted line corresponds to 
$R_e$, the dashed line corresponds to $R_\mu$ and the solid line corresponds to 
$R_\tau$ respectively.} 
\end{center}
\end{minipage}&\hspace{2cm}&
\begin{minipage}{0.4\hsize}
\begin{center}
\includegraphics[width=6cm]{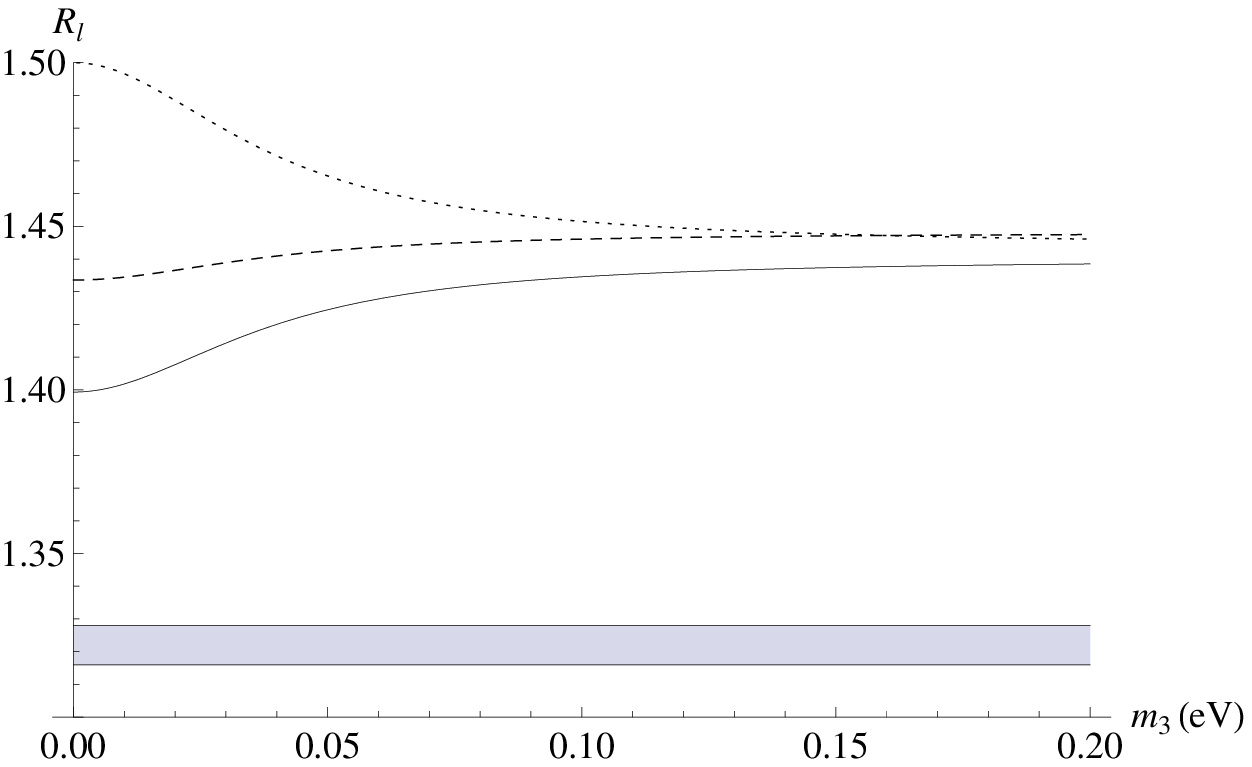}
\caption{$R_l$ $(l=e, \mu, \tau)$ for inverted hierarchical case as  functions 
of the lightest neutrino mass $m_3$ (eV).
The dotted line corresponds to 
$R_e$, the dashed line corresponds to $R_\mu$ and the solid line corresponds to $R_\tau$ respectively.}
\end{center}
\end{minipage}
\end{tabular}
\end{center}
\end{figure}

As we have seen from Fig.~8 and Fig.~9, we can expect the sizable deviation
from the standard model expectation on the a single charged lepton jet 
production rate with large missing momentum in 
the neutrinophilic two Higgs doublet model. The ratio $R_l$ can deviate
about $10 \%$ level.  The sizable flavor dependence comes from the following
reasons.
\noindent \\
(1) 
The non-suppressed charged Higgs and neutral Higgs production cross section
for non-degenerate charged and neutral Higgses as shown in Fig.~7. \\
(2) The flavor specific charged Higgs decay into chraged lepton  
strongly depends on two different patterns of hierarchy of neutrino mass spectrum
 corresponding to the normal or inverted.
It also depends on the Maki Nakagawa Sakata (MNS) matrix.\\
Therefore we conclude that $R_l$ is a key measurement which distinguishes the present 
model from the standard model.\\
\noindent
{\bf Acknowledgement}
T. M. was supported by KAKENHI, Grant-in-Aid for 
Scientific Research(C) No.22540283 from JSPS, Japan.
\bibliographystyle{unsrt}

\end{document}